# High-Capacity Framework for Reversible Data Hiding in Encrypted Image Using Pixel Prediction and Entropy Encoding

Yingqiang Qiu, Qichao Ying, Yuyan Yang, Huanqiang Zeng, *Senior Member, IEEE*, Sheng Li, *Member, IEEE*, and Zhenxing Qian, *Member, IEEE*

*Abstract*—While the existing vacating room before encryption (VRBE) based schemes can achieve decent embedding rate, the payloads of the existing vacating room after encryption (VRAE) based schemes are relatively low. To address this issue, this paper proposes a generalized framework for high-capacity RDHEI for both VRBE and VRAE cases. First, an efficient embedding room generation algorithm (ERGA) is designed to produce large embedding room by using pixel prediction and entropy encoding. Then, we propose two RDHEI schemes, one for VRBE, another for VRAE. In the VRBE scenario, the image owner generates the embedding room with ERGA and encrypts the preprocessed image by using the stream cipher with two encryption keys. Then, the data hider locates the embedding room and embeds the encrypted additional data. In the VRAE scenario, the cover image is encrypted by an improved block modulation and permutation encryption algorithm, where the spatial redundancy in the plain-text image is largely preserved. Then, the data hider applies ERGA on the encrypted image to generate the embedding room and conducts data embedding. For both schemes, the receivers with different authentication keys can respectively conduct error-free data extraction and/or error-free image recovery. The experimental results show that the two proposed schemes outperform many state-of-the-art RDHEI arts. Besides, the schemes can ensure high security level, where the original image can be hardly discovered from the encrypted version before and after data hiding by the unauthorized user.

*Index Terms*—Reversible data hiding, encrypted image, entropy encoding, image trust, copyright protection

## I. INTRODUCTION

Reversible data hiding (RDH) has the capability of accurately recovering the cover medium after the hidden data is extracted. For this reason, RDH is widely applied in some important and sensitive applications, i.e., covert communication, data hiding for medical images and law-enforcement, etc. Over the past two decades, many RDH researches have been done [1], most of which are developed for uncompressed images. Traditionally, RDH focuses on enlarging the embedding capacity and minimizing the distortion with the criterion of peak signal-to-noise ratio (PSNR). Typical RDH techniques can be classified into three categories, i.e., difference expansion (DE) [2-5], histogram shifting (HS) [6-9], and entropy coding [10-12]. In addition, there are many RDH schemes for JPEG images [13-15].

With the growing development of remote service, cloud security has attracted considerable attention. For data privacy protection, the encryption technique is widely used in cloud systems. In many applications, the cloud server needs to embed some additional data into the encrypted multimedia, such as copyright data, timestamp, etc. However, the spatial redundancy of cover multimedia is largely reduced during encryption. As a result, the traditional RDH methods are generally not efficient for the encrypted image. Therefore, in recent years, RDH in encrypted image (RDHEI) has attracted extensive research interest.

Fig.1 depicts the general framework of RDHEI. Typically, there are three parties involved in the data flow, i.e., the image owner, the data hider, and the receiver. The image owner firstly encrypts the cover image with or without some image preprocessing, i.e., whether the owner generates embedding room before image encryption. Then, he uploads the encrypted image onto the cloud. Next, the cloud server, also the data hider, embeds some encrypted additional data into the encrypted image. Finally, the receiver with different authentication keys can extract the embedded data error-free and/or recover the original image losslessly.

The existing RDHEI methods can be divided into two categories by different application scenarios [1]: vacating room after encryption (VRAE) [16-24] and vacating room before encryption (VRBE) [25-37]. The main difference lies in that whether the image owner vacates the embedding room ahead of image encryption by preprocessing or not.

Puech et al. proposes the first RDHEI scheme [16], in which the data hider vacates the embedding room within the encrypted version of the cover image. Since then, VRAE-based RDHEI schemes have received extensive attention. In [17], Zhang firstly encrypts the cover image using the stream cipher. Then,

This work was supported by the Natural Science Foundation of China (Grant U20B2051, 61972168, 62002124, 61871434). Corresponding author: Zhenxing Qian, Sheng Li.

Y. Qiu, Y. Yang and H. Zeng are with the College of Information Science & Engineering, Huaqiao University, Xiamen 361021, China, Y. Yang is also with Modern Engineering Training Base, Xiamen University of Technology, Xiamen 361021, China. (e-mail: yqqiu@hqu.edu.cn, yyy@xmut.edu.cn, zeng0043@hqu.edu.cn).
Q. Ying, S. Li and Z. Qian are with Shanghai Institute of Intelligent Electronics & Systems, School of Computer Science, Fudan University, Shanghai 200433, China. (e-mail: shinydotcom@163.com, lisheng@fudan.edu.cn, zxqian@fudan.edu.cn).

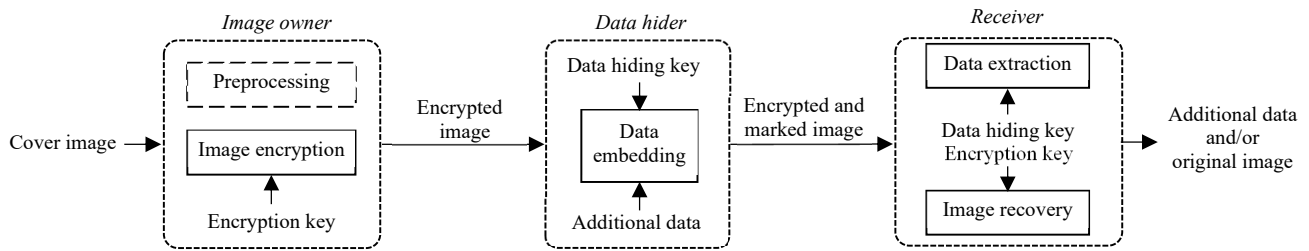

Fig. 1. General framework of RDHEI. Different from traditional RDH scenarios, the image owner encrypts the cover image to prevent information leakage and the data hider embeds his data in the encryption domain. The embedding room can be vacated either before or after the encryption. On the recipient's side, the extraction of the hidden data and the recovery of the original image can be separately conducted.

each selected block of the encrypted image can accommodate one bit by flipping or keeping three LSBs of all pixels in the block. The corresponding data extraction and image recovery are operated by observing the smoothness of each block. However, with the block size decreases, namely the amount of embedded data increases, extracted-bit errors may occur during data extraction and image recovery. To decrease the bit error rate, Hong et al. [18] improves this method by using a side match mechanism for measuring the smoothness of blocks. Zhou et al. [19] proposes to embed additional data in the encryption domain by using a public key modulation mechanism, and designs a two-class SVM classifier to distinguish the encrypted and nonencrypted image patches at the decoding side. In these VRAE-based methods [17-19], each image block can accommodate one bit, and therefore the achieved embedding capacity is relatively low. Besides, the operation of data extraction and image recovery should be done jointly after marked image decryption. But these schemes may also have errored bits during data extraction.

Some researchers have proposed separable VRAE-based RDHEI schemes. The first RDHEI scheme [16] is separable, where the embedded data is extracted before image decryption and the original image is recovered after decryption. In [20], Zhang further proposes a separable RDHEI method for stream-enciphered images, where the encrypted image is divided into blocks, and the LSBs of each block are compressed to generate embedding room for additional data. By using low density parity check codes (LDPC), Qian et al. [21] compresses the selected bits taken from the stream-ciphered image into syndrome bits, vacating room to accommodate some additional data. In [22], the original image is encrypted by public key cryptosystems with probabilistic and homomorphic properties. Base on the benefit of homomorphic encryption, RDH can be achieved in the image encryption domain. Since the entropy of an encrypted image is generally high, the achieved embedding capacities of these VRAE-based methods [16-22] are relatively low. In [23], Yi et al. proposes a separable VRAE-based RDHEI method with high embedding capacity. The image owner firstly divides the cover image into non-overlapping blocks and then encrypts it with block permutation and co-modulation (BPCM). The spatial redundancy of each image block is preserved well in a certain extent. After the encrypted image is uploaded to the cloud, the data hider can embed a large amount of additional data into the encrypted image using parametric binary tree labeling (PBTL).

Different from VRAE-based methods, VRBE-based methods allow the image owner to vacate embedding room ahead of image encryption. Therefore, the spatial redundancy in the original image can be exploited so that a higher embedding capacity can be achieved. Ma et al. [25] proposes the first VRBE-based RDHEI scheme based on histogram shifting. A considerable amount of vacated room can be reserved by the image owner and further utilized by the data hider. The work has received extensive attention and many improved schemes have been proposed. In [26], Zhang et al. proposes to embed additional data into the prediction errors of some selected pixels to generate embedding room before image encryption. In [27], the prediction errors generated by an interpolation technique are encrypted by a novel mode. In [28], the embedding room is vacated using patch-level sparse representation before image encryption. And RDH in homomorphic encrypted image vacates room is achieved by self-embedding in [29]. However, the embedding capacity is still relatively low.

To further increase the payload of RDHEI, Qiu et al. [30] performs block-wise adaptive reversible integer transformation (ARIT) on the cover image, and a large amount of embedding room can be vacated in the LSBs of the transformed image blocks. Similarly, Qiu et al. [31] proposes an RDHEI method based on ARIT, where high embedding capacity can be provided for both the image owner and the cloud server. Peneaux et al. proposes two RDHEI approaches [32] based on MSB prediction, i.e., high-capacity reversible data hiding approach with correction of prediction errors (CPE-HCRDH) and high-capacity reversible data hiding approach with embedded prediction errors (EPE-HCRDH). Base on the PBTL proposed in [23], Wu et al. [33] uses the redundancy of the entire plain-text image instead of the encrypted image blocks to vacate a larger embedding room before image encryption. Chen et al. [34] uses the correlations of high bit planes of plain-text image to create embedding room, which is improved by Yin et al. [35] using both pixel prediction and high bit planes compression of prediction errors to achieve a higher embedding capacity. Besides, Yin et al. [36] further proposes a high-capacity RDHEI scheme based on multi-MSB prediction and Huffman encoding. Compared with [25-29], [30-36] can achieve much higher embedding capacity.

Though both the VRAE-based and VRBE-based methods can provide promising embedding performances, there is still a discrepancy between them, many efficient embedding strategies proposed for one category cannot be well applied to the other. Previously, there are works that utilizes pixel prediction and entropy encoding [23,33]. However, the payload is not high especially for texture images, and payloads of the two methods vary a lot. Therefore, it is urgent to develop a framework that can work on both VRAE and VRBE scenarios with high embedding payload. Besides, there are some security flaws in current methods. For instance, Qu et al. [24] analyses the security of the RDHEI algorithm based on BPCM encryption [23] under known plaintext attacks (KPAs). This scheme shows that BPCM encryption has the risk of information leakage. The work in [33] indicates that a fingerprint of the cover image can be obtained from the PBTL in encryption domain, which corresponds to its prediction-error histogram.

To address the above issues, this paper proposes a generalized framework for high-capacity RDHEI for both VRBE and VRAE cases using pixel prediction and entropy encoding. We denote the proposed schemes as PE-VRBE and PE-VRAE, respectively. We begin with proposing an embedding room generation algorithm (ERGA) to produce embedding room. Firstly, we generate the prediction-error histogram (PEH) of the selected cover using the adjacency prediction and median edge detector (MED). According to the prediction errors, we divide the pixels into independent encoding pixels and joint encoding pixels. We then compress prediction errors with arithmetic coding and conduct self-embedding. In this way, high-capacity embedding room can be generated for RDHEI. In the VRBE scenario, the owner applies the proposed ERGA on the plain-text cover image to generate large embedding room. Then he encrypts the preprocessed image using the stream cipher, which avoids the content leakage of cover image. In the cloud, the data hider locates the vacated room and embeds the encrypted additional data. In the VRAE scenario, the image owner firstly encrypts the cover image by the improved block modulation and block permutation. The spatial redundancy in the plain-text image can be largely preserved within image blocks after image encryption. Afterwards, the data hider applies the proposed ERGA on the encrypted blocks to vacate room and embed his information. At the recipient side in both the VRAE and VRBE scenarios, error-free data extraction and lossless image recovery can be separably conducted in a reverse way, i.e., the receivers with different authentication keys get different results.

Experimental results demonstrate the effectiveness and superiority of the proposed schemes compared to previous arts. The proposed schemes achieve much higher embedding capacity than many state-of-the-art RDHEI methods. At the same time, the proposed schemes provide high information security. Interestingly, although the payloads of previous VRAE-based schemes are reported to be naturally lower than those of VRBE-based schemes, we find that our PE-VRAE can beat some state-of-the-art VRBE-based schemes, owing to the effectiveness of the redundancy preserving room reservation algorithm. Besides, the payload of the proposed PE-VRAE scheme is close to that of the proposed PE-VRBE scheme. We also validate that the proposed schemes are effective in extreme cases, i.e., using textured images as cover images.

The main contributions of this paper are three-fold:
- We are the first to propose an RDHEI framework that can work on both the VRAE and VRBE cases.
- The proposed schemes under the framework are high in payload, which can beat state-of-the-art RDHEI schemes.
- Little information of the original image can be discovered from the encrypted image or the marked encrypted image by the unauthorized user. Besides, the proposed framework can provide large payload for the textured image.

The remainder of this paper is organized as follows: the proposed embedding room generation algorithm is detailed in Section II. The proposed VRBE-based scheme and VRAE-based scheme are then introduced in Section III and Section IV, respectively. Experimental results and discussions are provided in Section V. Section VI concludes this paper.

## II. Efficient Embedding Room Generation Algorithm

To construct a general RDHEI framework for both VRAE and VRBE, we begin with introducing a novel efficient embedding room generation algorithm (ERGA). We denote the targeted image before embedding room generation as **Y**, which is the plain-text image in VRBE and the encrypted image in VRAE. Without loss of generality, we assume that **Y** is an 8-bits gray-scaled image with a size of $N_1 \times N_2$. The proposed algorithm is composed of two steps: generation of the prediction-error histogram, and embedding room generation by optimal entropy encoding. In addition, we discuss the image recovery procedures through entropy decoding. The key notations used in this paper are summarized in Table 1.

Table 1. Summary of key notations in this paper

| Notation | Description |
|---|---|
| **X** | Cover image |
| **Y** | Targeted image before embedding room generation (cover image in VRBE, encrypted image in VRAE) |
| **Y<sub>R</sub>** | Encrypted image with vacated room |
| **Z** | Marked encrypted image |
| $Y_i$ | Image blocks of **Y** |
| $P_R$ | Reference pixels |
| $P_E$ | Embedding pixels |
| $P_I$ | Independent encoding pixels |
| $P_J$ | Joint encoding pixels |
| $B_I$ | Independent encoding bins |
| $B_J$ | Joint encoding bins |
| $py$ | Prediction value of pixel $y$ |
| $e$ | Prediction error |
| $h$ | Prediction-error histogram |
| $T$ | Threshold |
| $T_{opt}$ | Optimal threshold |
| $CD$ | Compressed data |
| $AD$ | Auxiliary data |
| $L$ | Total length of $CD$ and $AD$ |
| $EC_T$ | Embedding capacity under $T$ |

### A. Prediction Errors Generation

Prediction error generation aims at compressing the bit-length for representing the original pixels in the selected cover image. The technique is widely applied in RDH-based schemes [23, 33] to vacate embedding room by utilizing the spatial redundancy. Apart from using the traditional MED predictor, we further make predictions on the pixels using adjacency prediction for larger amount of embedding room.

Table 2. Pixel assignment of the MED predictor for image block sized $n=4\times 4$

| $a$ | $b$ | $c$ | $y$ | $a$ | $b$ | $c$ | $y$ | $a$ | $b$ | $c$ | $y$ |
|---|---|---|---|---|---|---|---|---|---|---|---|
| $y_{i,2,2}$ | $y_{i,2,1}$ | $y_{i,1,2}$ | $y_{i,1,1}$ | $y_{i,2,2}$ | $y_{i,2,3}$ | $y_{i,1,2}$ | $y_{i,1,3}$ | $y_{i,2,3}$ | $y_{i,2,4}$ | $y_{i,1,3}$ | $y_{i,1,4}$ |
| $y_{i,2,2}$ | $y_{i,2,1}$ | $y_{i,3,2}$ | $y_{i,3,1}$ | $y_{i,2,2}$ | $y_{i,2,3}$ | $y_{i,3,2}$ | $y_{i,3,3}$ | $y_{i,2,3}$ | $y_{i,2,4}$ | $y_{i,3,3}$ | $y_{i,3,4}$ |
| $y_{i,3,2}$ | $y_{i,3,1}$ | $y_{i,4,2}$ | $y_{i,4,1}$ | $y_{i,3,2}$ | $y_{i,3,3}$ | $y_{i,4,2}$ | $y_{i,4,3}$ | $y_{i,3,3}$ | $y_{i,3,4}$ | $y_{i,4,3}$ | $y_{i,4,4}$ |

We assume that $Y_i$ is the non-overlapping image block sized $n = n_1 \times n_2$, and the number of blocks is $N = \lfloor N_1/n_1 \rfloor \times \lfloor N_2/n_2 \rfloor$, where $\lfloor \cdot \rfloor$ stands for the floor function, $1 \leq i \leq N$. For each block $Y_i$, we pseudo-randomly select one pixel $y_{i,j_1,k_1}$ ($1 \leq j_1 \leq n_1, 1 \leq k_1 \leq n_2$) as the ***reference pixel*** ($P_R$), whose value is unchanged during the embedding room generation. The seed of the pseudo-random sequence generator is fixed or shared with the data hider and the recipients. The other pixels are served as the ***embedding pixels*** ($P_E$), which are predicted by the adjacent pixels for embedding room generation. The prediction of $P_E$ is conducted as follows.

Case1: pixels in the horizontal directions of $y_{i,j_1,k_1}$,

$$py_{i,j_1,k} = \begin{cases} y_{i,j_1,k+1}, & 1 \leq k < k_1 \\ y_{i,j_1,k-1}, & k_1 < k \leq n_2 \end{cases}. \quad (1)$$

Case2: pixels in the vertical directions of $y_{i,j_1,k_1}$,

$$py_{i,j,k_1} = \begin{cases} y_{i,j+1,k_1}, & 1 \leq j < j_1 \\ y_{i,j-1,k_1}, & j_1 < j \leq n_1 \end{cases}. \quad (2)$$

Case3: other pixels in the upper left, upper right, bottom left or bottom right side of $y_{i,j_1,k_1}$ are predicted by the MED predictor. Accordingly, four MED prediction templates are shown in Fig. 2, where $y$ represents the pixel to be predicted, and the three neighboring pixels of $y$ are denoted by $a$, $b$, and $c$ respectively. The prediction value $py$ of $y$ is calculated by:

$$py = \begin{cases} \max(b,c), & a \leq \min(b,c) \\ \min(b,c), & a \geq \max(b,c) \\ b + c - a, & \text{otherwise} \end{cases} \quad (3)$$

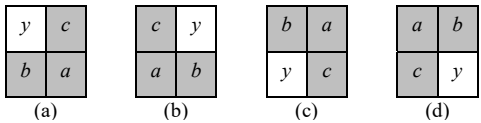

Fig. 2. Four prediction templates for MED. In each prediction mode, $y$ is predicted by $a$, $b$ and $c$ with Eq. (3).

In this way, $(n − 1)$ prediction values can be obtained in each image block. Fig. 3 shows an example of prediction error generation for an image block sized $n = 4 \times 4$. We regard $y_{i,2,2}$ as the reference pixel, the other pixels are predicted by the following operation.

Case1: on the horizontal direction, $y_{i,2,1}$ and $y_{i,2,3}$ are predicted with $y_{i,2,2}$, $y_{i,2,4}$ is predicted with $y_{i,2,3}$;

Case2: on the vertical direction, $y_{i,1,2}$ and $y_{i,3,2}$ are predicted with $y_{i,2,2}$, $y_{i,4,2}$ is predicted with $y_{i,3,2}$;

Case3: on the corners, the pixel assignments of $a$, $b$, $c$, and $y$ are shown in Table 2, and $y$ is predicted with Eq. (3).

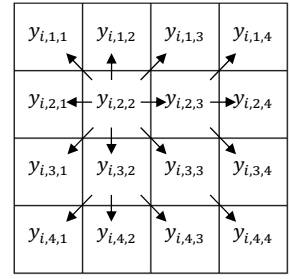

Fig. 3. Example of pixel prediction for an image block sized $n=4\times4$. Arrows show the prediction direction.

Afterwards, the prediction error $e_{i,j,k}$ is calculated by

$$e_{i,j,k} = y_{i,j,k} - py_{i,j,k}, \quad (4)$$

where $e_{i,j,k} \in [-255, 255]$, $j \neq j_1$ or $k \neq k_1$. Hence, $N \cdot (n-1)$ prediction errors are obtained.

After calculating all prediction errors, we construct a prediction-error histogram (PEH) represented as $h$ by

$$h = \{h_e | h_e = \#(e_{i,j,k} = e), e \in [-255, 255]\}, \quad (5)$$

where $\#$ denotes the amount of satisfied prediction errors. The PEH is denoted as $h = \{h_{-255}, h_{-254}, \ldots, h_{-1}, h_0, \ldots, h_{254}, h_{255}\}$, where the reference pixels are excluded.

### B. Room Generation by Optimized Arithmetic Encoding

For a plain-text image in VRBE, the PEH generally obeys a Laplacian-like distribution centered at zero. The value of $h_e$ decreases along with the absolute value of $e$ increases. Fig. 4 shows an example of the PEH of the plain-text image Lena.

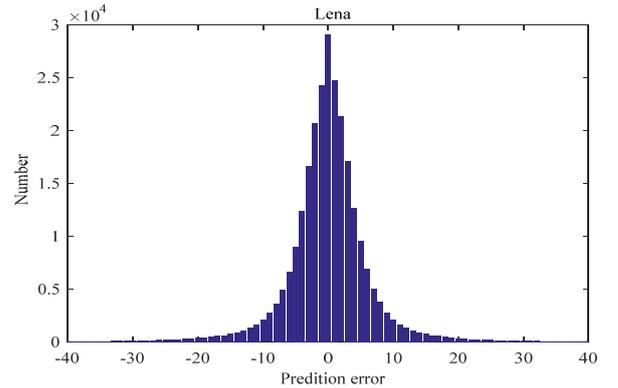

Fig. 4. The prediction-error histogram of Lena, which obeys a Laplacian-like distribution.

We separate the histogram bins $\{h_{-255}, h_{-254}, \ldots, h_{-1}, h_0, \ldots, h_{254}, h_{255}\}$ of the PEH into two categories, namely, the ***independent encoding bins*** $B_I = \{h_{-T}, h_{-T+1}, \ldots, h_{-1}, h_0, \ldots, h_{T-2}, h_{T-1}\}$ and the ***joint encoding bins*** $B_J = \{h_{-255}, h_{-254}, \ldots, h_{-T-2}, h_{-T-1}, h_T, h_{T+1}, \ldots, h_{254}, h_{255}\}$, $h = B_I \cup B_J$. $T$ is an adaptive threshold that alters with the PEH, i.e., different cover images use different $T$ for highest payload. Accordingly, we define the embedding pixels as *independent encoding pixels* ($P_I$) or *joint encoding pixels* ($P_J$). In this way, all prediction errors are classified into ($2T+1$) categories, which can be compressed losslessly with traditional entropy-coding methods, and the length of compressed data (*CD*) can be denoted as $\ell_T(CD)$. Since the pixels in $P_J$ are jointly encoded, their original values are kept as auxiliary data (*AD*) for image recovery. The length is

$$\ell_T(AD) = 8h_J, \quad (6)$$

where $h_J = \sum_{e=-255}^{-T-1} h_e + \sum_{e=T}^{255} h_e$. The total length $L$ of *CD* and *AD* must satisfy

$$L = \ell_T(CD) + \ell_T(AD) < 8N(n_1 n_2 - 1) < 8N_1 N_2, \quad (7)$$

which needs to be recorded with $\ell(L)$ bits, and

$$\ell(L) = 3 + \lceil \log_2(N_1 N_2) \rceil, \quad (8)$$

where $\lceil \cdot \rceil$ is the ceiling function. Therefore, for every distinctive $T$, the net payload of vacated room is

$$EC_T = 8 \cdot \sum_{e=-255}^{255} h_e - \ell_T(CD) - \ell_T(AD) - \ell(L)$$
$$= 8 \cdot \sum_{e=-T}^{T-1} h_e - \ell_T(CD) - \ell(L). \quad (9)$$

Afterwards, the target is to find the optimized $T \in [1, 255]$ with the largest amount of vacated room as $T_{opt}$. Arithmetic code and Huffman code are the two common entropy encoding paradigms. Here we take arithmetic encoding as an example.

For a given $T \in [1, 255]$, the *independent encoding pixels* with the same prediction error $e \in [-T, T-1]$ are assigned with an independent symbol, all *jointly encoding pixels* are assigned with a same symbol. Therefore, the numbers of ($2T+1$) symbols are $\{h_{-T}, h_{-T+1}, \ldots, h_{-1}, h_0, \ldots, h_{T-2}, h_{T-1}, h_J\}$ respectively, which are recorded for arithmetic decoding. Since it is less than the number of pixels $N_1 \cdot N_2$, the number of each symbol can be expressed within $\lceil \log_2(N_1 N_2) \rceil$ bits. The length of output data compressed with arithmetic coding is expressed with $(3 + \lceil \log_2(N_1 N_2) \rceil)$ bits. For $T \in [1, 255]$, it can be expressed within eight bits. Thus, the above side information is concatenated as a binary sequence $CD_1$ with length $\ell_T(CD_1)$ as

$$\ell_T(CD_1) = 11 + (2T+2) \cdot (\lceil \log_2(N_1 N_2) \rceil). \quad (10)$$

Then, the occurrence of each symbol for $e \in [-T, T-1]$ can be calculated as

$$p_e = h_e / \sum_{e'=-255}^{255} h_{e'}. \quad (11)$$

The total occurrence of the symbol for $P_J$ is

$$p_J = 1 - \sum_{e=-T}^{T-1} p_e. \quad (12)$$

With $\{p_{-T}, p_{-T+1}, \ldots, p_{-1}, p_0, \ldots, p_{T-2}, p_{T-1}, p_J\}$, all symbols assigned for the *embedding pixels* can be compressed entirely to be a binary sequence $CD_2$ with length $\ell_T(CD_2)$. Then, $CD_1$ is appended with $CD_2$, and the length is

$$\ell_T(CD) = \ell_T(CD_1) + \ell_T(CD_2) = 11 + \ell_T(CD'), \quad (13)$$

where $CD'$ is a part of $CD$, whose length is

$$\ell_T(CD') = (2T+2) \cdot \lceil \log_2(N_1 N_2) \rceil + \ell_T(CD_2). \quad (14)$$

With arithmetic coding, the net payload of vacated room in Eq. (9) can be described as

$$EC_T = 8 \cdot \sum_{e=-T}^{T-1} h_e - 11 - \ell(L) - \ell_T(CD'). \quad (15)$$

Then, to achieve maximal embedding capacity, the optimal threshold $T_{opt}$ should satisfy

$$T_{opt} = \underset{T \in [1,255]}{\arg\max} EC_T. \quad (16)$$

With arithmetic coding, we can compress the *embedding pixels* of image **Y** to be *CD* and *AD*. We skip all reference pixels and replace the first $\ell(L)$ bits in the LSB plane of all embedding pixels with the value of $L$. Subsequently, the lower bit planes of all embedding pixels are replaced by the binary sequence of *CD* and *AD*, which can be encrypted if needed. The other higher bit planes of all embedding pixels are vacated for data embedding in the image encryption domain, and $L$ indicates to the authorized data hider or receiver the location of embedding room. The resulted image **Y'** is the image with vacated room.

*C. Pixel Recovery Using Entropy Decoding*

In this part, we discuss how to recover the pixels in **Y** from the image with vacating room **Y'**. We locate the referenced pixels using the shared pseudo-random sequence, and skip all the referenced pixels together with the first $\ell(L)$ bits in the LSB plane of all embedding pixels. We obtain $T_{opt}$ value from the next 8 bits in the lower bit planes.

With $T_{opt}$, the numbers of ($2T+1$) symbols $\{h_{-T}, h_{-T+1}, \ldots, h_{-1}, h_0, \ldots, h_{T-2}, h_{T-1}, h_J\}$, can be determined by every following $\lceil \log_2(N_1 N_2) \rceil$ bits in the lower bit planes of all embedding pixels. The length $\ell(CD_2)$ is determined by the next $(3 + \lceil \log_2(N_1 N_2) \rceil)$ bits. We then take the following $\ell(CD_2)$ as the compressed data stream $CD_2$. After calculating $\{p_{-T}, p_{-T+1}, \ldots, p_{T-2}, p_{T-1}, p_J\}$ with $\{h_{-T}, h_{-T+1}, \ldots, h_{T-2}, h_{T-1}, h_J\}$, we obtain the prediction errors of all embedding pixels by arithmetic decoding. Then, with the determined $h_J$, the auxiliary data *AD* is recovered.

After entropy decoding, all *joint encoding pixels* $P_J$ can be recovered error-free with the extracted *AD*. For each $y_{i,j,k} \in P_I$ of each image block, the corresponding prediction error $e_{i,j,k}$ is identified through entropy decoding. With $P_R$, recovered $P_J$ and $P_I$, the prediction value $py_{i,j,k}$ can be obtained pixel by pixel by using Eq. (1)−(3). Afterwards, the original value $y_{i,j,k}$ is recovered with Eq. (17),

$$y_{i,j,k} = e_{i,j,k} + py_{i,j,k}. \quad (17)$$

The recovered $P_I$ is used to predict the subsequent unrecovered $P_I$. In this way, the original pixel values of all $P_I$ are recovered iteratively. As a result, the image **Y** is recovered error-free.

III. THE PROPOSED VRBE-BASED RDHEI SCHEME

Fig. 5 depicts the architecture of the proposed VRBE-based RDHEI scheme (PE-VRBE). The scheme contains three parts:
1) *the owner side*, where the embedding room is vacated in

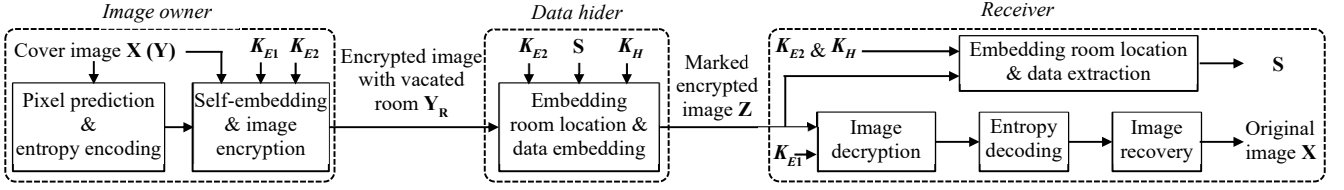

Fig. 5. The frameworks of the proposed PE-VRBE RDHEI scheme. We propose a novel ERGA for pixel prediction so that image owner can vacate a large embedding room using entropy encoding. Data extraction and image recovery are the inverse of the embedding procedures.

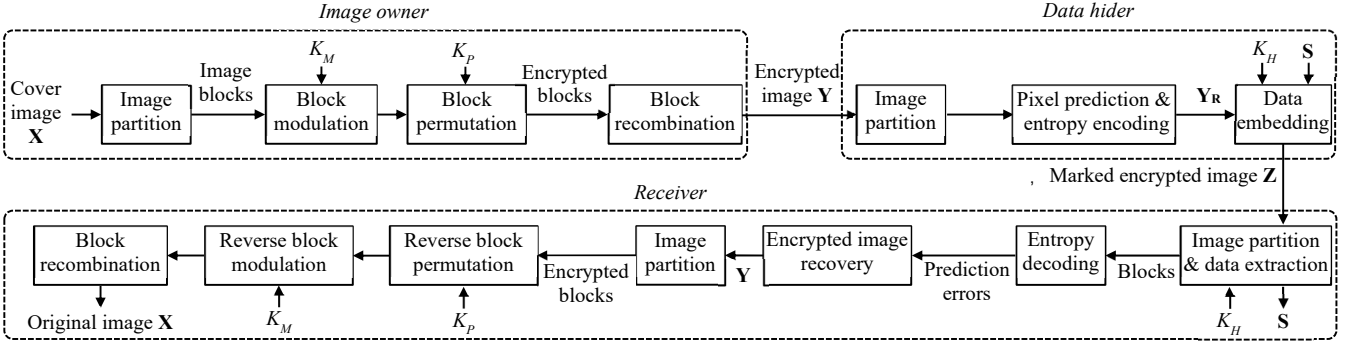

Fig. 6. The frameworks of the proposed PE-VRAE RDHEI scheme. The image owner encrypts the cover image with the improved block modulation and permutation, where spatial redundancy is largely preserved after image encryption. The data hider is allowed to generate a large embedding room even in the encryption domain. Data extraction and image recovery are separable.

the original image **X** (denoted as **Y** in Section II) by self-embedding, and the encrypted image with vacated room **Y_R** is produced. 2) *the data hider side*, where the additional data **S** is encrypted and embedded into **Y_R**, and 3) *the receiver side*, where data extraction and image recovery are separately done by the authorized users.

*A. Image Encryption with Vacating Room*

At the owner side, the image owner takes **Y** as an image block, and selects one pixel of **Y** as the reference pixel, for example, the pixel in the first row and the first column of **Y**. Then, he uses the adjacency prediction and MED predictor discussed in Section II to generate the PEH.

Then, with arithmetic coding, the owner finds out the optimal threshold $T_{opt}$. He compresses the prediction errors losslessly to obtain the compressed data *CD* and auxiliary data *AD*, and calculates $L = \ell(CD) + \ell(AD)$ with length $\ell(L)$. Afterwards, he skips the reference pixel and the first $\ell(L)$ bits in the LSB plane, and replaces the lower bit planes with the *CD* and *AD*. As a result, the rest bit planes are vacated as the embedding room. The processed image is denoted as **V**.

After the above room generation procedures, the owner encrypts **V** to obtain the encrypted version **V'** using the stream cipher. With modulo-2 operation in Eq. (18), a gray value $v_{j,k}$ ($1 \leq j \leq N_1, 1 \leq k \leq N_2$) in **V** can be represented by eight bits as $v_{j,k,l}$ ($0 \leq l \leq 7$).

$$v_{j,k,l} = \lfloor v_{j,k}/2^l \rfloor \bmod 2. \quad (18)$$

The owner generates a pseudo-random sequence *P* with the same size of the cover image according to Eq. (19). The encryption key $K_{E1}$ is used as the seed of the pseudo-random generator.

$$P = \{p_{j,k} | p_{j,k} \in [0,255], 1 \leq j \leq N_1, 1 \leq k \leq N_2\}. \quad (19)$$

Each $p_{j,k}$ is also represented by eight bits as $p_{j,k,l}$ as follow.

$$p_{j,k,l} = \lfloor p_{j,k}/2^l \rfloor \bmod 2. \quad (20)$$

Then, the image owner uses the bit-wise exclusive-or (XOR) function to encrypt each bit $v_{j,k,l}$ as

$$v'_{j,k,l} = v_{j,k,l} \oplus p_{j,k,l}, \quad (21)$$

where "⊕" is the XOR function, and $v'_{j,k,l}$ is the encrypted bit. Then, the encrypted pixel $v'_{j,k}$ can be calculated as

$$v'_{j,k} = \sum_{l=0}^{7} v'_{j,k,l} \cdot 2^l. \quad (22)$$

Therefore, the encrypted image **V'** is generated by the stream cipher, where $K_{E1}$ is required as the image encryption key.

With the encryption key $K_{E2}$ and stream cipher, the total length *L* of *CD* and *AD* can be encrypted to obtain *L'*. $K_{E2}$ should be distributed to the authorized data hider and receiver. Otherwise, the embedding room cannot be located by the recipients. The owner embeds *L'* into **V'** by replacing the first $\ell(L)$ bits in the LSB plane of **V'** to obtain the encrypted image with vacated room **Y_R**.

*B. Data Embedding*

The cloud server tempts to hide his data into the processed image. On receiving **Y_R**, the data hider locates the embedding room with the help of *L'*. He skips the reference pixel and extracts the first $\ell(L)$ bits in the LSB plane of **Y_R**, and then decrypts the obtained bits with $K_{E2}$. As a result, the length *L* of *CD* and *AD* is determined.

Afterwards, the additional data **S** to be embedded into **Y_R** is encrypted with the data hiding key $K_H$ to obtain **S'**. With *L*, the hider skips the lower bit planes carrying *CD* and *AD*, and replaces the higher bit planes with **S'**. In this way, the marked encrypted image **Z** based on VRBE is generated.

## C. Data Extraction and Image Recovery

At the recipient side, the process of data extraction and image recovery can be conducted separately. The authorized receiver extracts the embedded data with $K_{E2}$ and $K_H$, and/or recovers the original image with $K_{E1}$ and $K_{E2}$ respectively.

The receiver first locates the embedding room under the same principle of that used by the data hider. He skips the reference pixel and extracts the first $\ell(L)$ bits in the LSB plane of **Z**. Afterwards, he decrypts the extracted data with $K_{E2}$ to obtain the value of **L**. He further skips the reference pixel and the lower bit planes carrying *CD* and *AD* with **L** to extract the embedded data **S'** from the higher bit planes. Finally, the receiver with $K_H$ decrypts **S'** to obtain the additional data **S** error-free.

Since *CD* and *AD* are encrypted with $K_{E1}$. With $K_{E1}$, the receiver decrypts **Z** to **Z'**. As a result, the *CD* and *AD* can be recovered error-free in the lower bit planes of **Z'**, and the reference pixel is recovered exactly. Then, according to Section II-C, the receiver recovers the original image **Y**, namely **X** without error.

If the receiver has $K_{E2}$ and $K_H$ alone, he is only capable of extracting the embedded data **S** error-free. If the receiver has $K_{E1}$ and $K_{E2}$ alone, he can recover the original image losslessly. With all above three keys simultaneously, he is capable to not only extract **S** but also recover the cover image without error.

## IV. THE PROPOSED VRAE-BASED RDHEI SCHEME

Fig. 6 depicts an overview of the proposed VRAE-based RDHEI scheme (PE-VRAE). The scheme is also consisted of three phases: 1) the image owner side, where the cover image **X** is encrypted as **Y** by block modulation and permutation, 2) the data hider side, where the additional data **S** is embedded into **Y** to generate the marked encrypted image **Z**, and 3) the receiver side, where the embedded data can be extracted error-free and/or the original image can be recovered losslessly.

### A. Image Encryption with Correlation Preservation

At the first phase, the image owner generates the encrypted image, while preserving the original pixel correlation within each image block for VRAE. The owner firstly divides the cover image **X** of size $N_1 \times N_2$ into non-overlapping image blocks sized $n = n_1 \times n_2$ as

$$X_i = (x_{i,1,1}, x_{i,1,2}, \cdots, x_{i,2,1}, x_{i,2,2}, \cdots, x_{i,n_1,n_2-1}, x_{i,n_1,n_2}), \quad (23)$$

where $1 \leq i \leq N, 1 \leq j \leq n_1, 1 \leq k \leq n_2$, and the number of blocks is $N = \lfloor N_1/n_1 \rfloor \times \lfloor N_2/n_2 \rfloor$.

The owner then encrypts the image blocks using an improved pixel modulation algorithm, which preserves the pixel correlation. In the previous scheme [23], pixel modulation is conducted by firstly generating a random data sequence $R = \{r_1, r_2, \cdots, r_N\}$ of size $N$ with an encryption key $K_M$ and then modulating the pixels with Eq. (24).

$$x'_{i,j,k} = (x_{i,j,k} + r_i) \bmod 256. \quad (24)$$

Let $\min(X_i)$ and $\max(X_i)$ denote the minimum and maximum values of $X_i$, respectively. When $r_i \in [0, 255 - \max(X_i)] \cup (255 - \min(X_i), 255]$, the difference value of any two pixels in $X_i$ remains unchanged after the modulation.

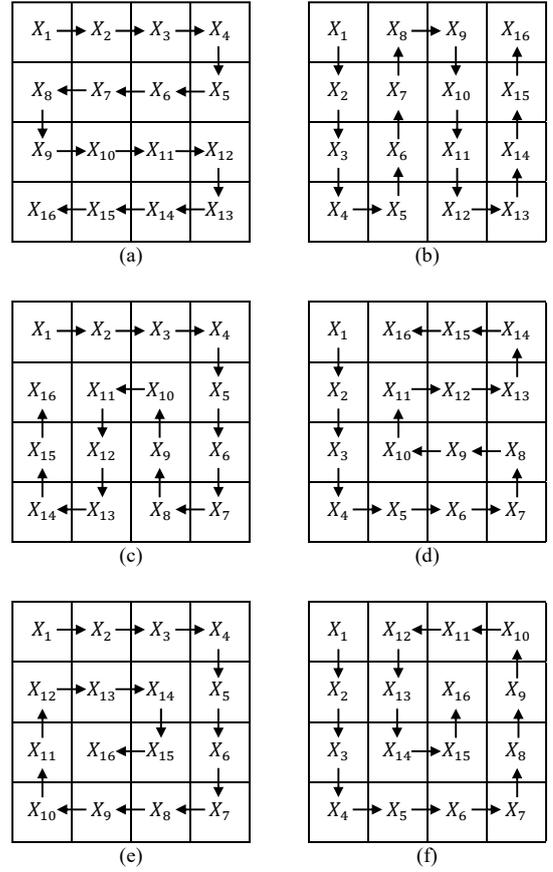

Fig. 7. Showcase of acceptable zig-zag ordering of the image blocks for pixel correlation preservation during block modulation ($N = 4 \times 4$).

$$x'_{i,j_1,k_1} - x'_{i,j_2,k_2} = x_{i,j_1,k_1} - x_{i,j_2,k_2}, j_1 \neq j_2 \text{ or } k_1 \neq k_2. \quad (25)$$

On the contrary, when $r_i \in (255 - \max(X_i), 255 - \min(X_i)]$, the pixel correlation of $X_i$ will be affected during the pixel modulation. Take $X_i = (202, 171, 166, 130)$, $r_i = 85$ as an example where $\min(X_i) = 130$ and $\max(X_i) = 202$, the spatial correlation will be influenced when $r_i \in (53, 125]$. After pixel modulation, the result is $X_i' = (31, 0, 251, 215)$. In this case, some differences between the adjacent pixels are changed by pixel modulation.

To decrease the influence of pixel modulation on the spatial correlation, we design an improved pixel modulation algorithm based on the adjacent image block. Fig. 7 shows some possible arrangements of image blocks where $N = 4 \times 4$, any ordering with subsequent blocks adjacent is acceptable. Since there is a spatial relationship between two adjacent blocks, a previous block $X_{i-1}$ is used to predict the ideal range of random data $r_i$ for the pixel modulation of $X_i$. According to $\min(X_{i-1})$ and $\max(X_{i-1})$, the ideal range of $r_{i-1}$ is $[0, 255 - \max(X_{i-1})] \cup (255 - \min(X_{i-1}), 255]$. The owner uses a scale factor as $\zeta \in [0,1]$ to limit the range as $[0, \zeta \cdot (255 - \max(X_{i-1}))] \cup (255 - \zeta \cdot \min(X_{i-1}), 255]$. Then, using $r_i$ as the seed, the owner generates a new random integer $r'_i \in [0, \zeta \cdot (255 - \max(X_{i-1}))] \cup (255 - \zeta \cdot \min(X_{i-1}), 255]$ for the pixel modulation of each image block $X_i$ except the first one $X_1$, where the owner directly sets $r'_1 = r_1$. Thus, the traditional pixel modulation is updated in Eq. (26), and the random data sequence $R' = \{r'_1, r'_2, \cdots, r'_N\}$ is generated using the encryption key $K_M$.

$$x'_{i,j,k} = (x_{i,j,k} + r'_i) \bmod 256. \tag{26}$$

The owner further permutes the modulated image blocks $X'_i$ using the Arnold transformation, which is described as:

$$\begin{bmatrix} k' \\ l' \end{bmatrix} = \begin{bmatrix} 1 & b \\ a & ab+1 \end{bmatrix} \begin{bmatrix} k \\ l \end{bmatrix} \bmod(M), \tag{27}$$

where $k$ and $l$ are the row and column of the image block before transformation, respectively, $k'$ and $l'$ are the row and column after transformation, $a$ and $b$ are the transformation parameter, and $M$ is the same number of image blocks on the row and column. If $\lfloor N_1/n_1 \rfloor \neq \lfloor N_2/n_2 \rfloor$, $M$ should be set to be the greatest common factor of $\lfloor N_1/n_1 \rfloor$ and $\lfloor N_2/n_2 \rfloor$. The parameters $a, b$ and $M$ are determined by the encryption key $K_P$. To resist the known plaintext attacks [24], the image encryption should be one-time pad.

In this way, the encrypted image with correlation preservation **Y** can be generated and uploaded to the cloud.

*B. Data Embedding in the Cloud*

At the second phase, the data hider embeds additional data into the encrypted image. The procedure of data embedding is composed of three steps: prediction errors generation, entropy encoding of prediction errors, and data embedding.

First, the data hider divides **Y** into non-overlapping blocks under the same principle of that in Section IV-A. For each block $Y_i$, a reference pixel $P_R$ denoted as $y_{i,j_1,k_1}$ is selected and the other pixels are served as $P_E$. Then, the data hider predicts all $P_E$ with the adjacency prediction and MED predictor, and generates the corresponding PEH.

Second, with arithmetic coding, the hider calculates the optimal threshold $T_{opt}$ and compresses the prediction errors losslessly to obtain the compressed data *CD* and auxiliary data *AD*. Then, he calculates $L = \ell(CD) + \ell(AD)$ and transfers it into $\ell(L)$ bits binary data. Next, the hider encrypts the additional data **S** with the data hiding key $K_H$ to obtain **S'**. The data to be embedded is composited of *L*, *CD*, *AD* and **S'**.

Finally, the hider skips the reference pixels and replaces each bit planes iteratively with the data to be embedded from LSB-plane to MSB-plane. As a result, the marked encrypted image **Z** is generated.

*C. Data Extraction and Image Recovery*

The data extraction and image recovery are separable, and the operation of data extraction is similar to that of PE-VRBE. The receiver firstly skips the reference pixels and extracts the first $\ell(L)$ bits in the LSB plane of **Z**, which is the length ***L*** of *CD* and *AD*. Then, with ***L***, the receiver skips the reference pixels and the lower bit planes carrying *CD* and *AD***,** and extracts the embedded data **S'** from the higher bit planes. Finally, the authorized receiver with $K_H$ decrypts **S'** to obtain **S** without error.

Thereafter, according to Section II-C, the receiver recovers the encrypted image **Y** without error. Without the encryption keys $K_P$ and $K_M$, little information of the original image can be discovered from **Y**. With $K_P$ and $K_M$, the receiver further recovers the original image. He divides **Y** into non-overlapping image blocks $\{Y_1, Y_2, \cdots, Y_N\}$ under the same principle of that in Section IV-A. With $K_P$ and Eq. (28), the receiver operates reverse permutation of blocks $\{Y_1, Y_2, \cdots, Y_N\}$ to obtain the modulated image blocks $\{X'_1, X'_2, \cdots, X'_N\}$.

$$\begin{bmatrix} k \\ l \end{bmatrix} = \begin{bmatrix} ab+1 & -a \\ -b & 1 \end{bmatrix} \begin{bmatrix} k' \\ l' \end{bmatrix} \bmod(M). \tag{28}$$

Next, he conducts the reverse pixel modulation with Eq. (29).

$$x_{i,j,k} = (x'_{i,j,k} - r'_i) \bmod 256. \tag{29}$$

The reverse modulation is carried out in the order of $\{X'_1, X'_2, \cdots, X'_{N-1}, X'_N\}$. Specifically, the receiver generates the random data sequence $R = \{r_1, r_2, \cdots, r_N\}$ with $K_M$. Then, with $r'_1 = r_1$, $X_1$ is recovered from $X'_1$ with Eq. (29). With $r_2$ and $X_1$, $r'_2$ is generated to recover $X'_2$. In this way, with recovered $X_{i-1}$ and $r_i$, the subsequent block $X_i (i > 2)$ is iteratively recovered. Finally, the original image **X** is recovered without error.

When the receiver has both the data hiding key and encryption keys, he is capable to not only extract the secret data exactly but also recover the cover image losslessly.

V. EXPERIMENTAL RESULTS

To verify the two RDHEI schemes under the proposed framework, we have conducted a series of experiments on thousands of gray-scaled images from typical image datasets, namely, UCID [38], BOSSBase [39], and BOWS-2 [40]. We use a binary random sequence as the additional data, where the possibilities of 0 and 1 are equal. In this section, we compare the net payload among different methods in terms of embedding rate (ER). The net payload is computed by excluding the side information which contains the compressed data (*CD*) and auxiliary data (*AD*). For objective image quality evaluation, we employ the peak signal-to-noise ratio (PSNR) and structural similarity (SSIM) [41] as evaluation metrics. A larger score of PSNR or SSIM indicates a better content similarity between two images.

*A. Validation of Redundancy Preservation in PE-VRAE.*

In PE-VRAE, the spatial correlation within each block can be influenced by the pixel modulation, which is related to the blocks size $n = n_1 \times n_2$ and the scale factor $\zeta$. The image owner generates embedding room with a random data, there are two ideal cases, one is that all processed pixel values are within [0, 255], the other is that all are larger than the ceiling value 255. However, in some cases, some processed pixel values may be within [0, 255], others are larger than 255. We regard these pixels, with less amount of which are no more than or larger than 255, as abnormal pixels. The total amount of all abnormal pixels reflects the correlation preservation within the blocks. Table 3 shows the experimental results for the five test images in Fig. 8 with different parameters $n = \{4 \times 4, 6 \times 6, 8 \times 8\}$ and ζ = {0.25, 0.50, 0.75, 1.00, none}, where "none" represents that we apply the traditional pixel modulation in [23] instead of our improved strategy. With the given block size, due to the interval expansion of random data for modulation, the amount of abnormal pixels increases with a larger ζ, while that amount is much larger when ζ= none. The test proves that our proposed pixel modulation strategy can largely preserve the spatial correlation within image blocks, which plays a crucial role in ensuring a larger payload.

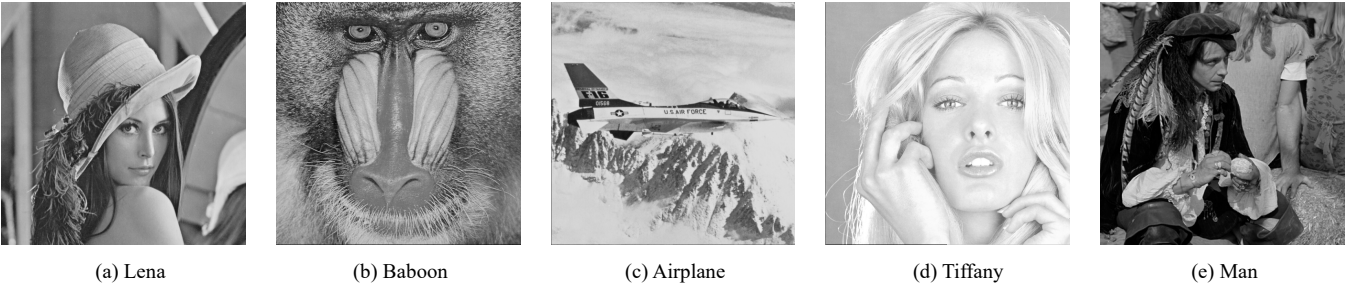

(a) Lena  (b) Baboon  (c) Airplane  (d) Tiffany  (e) Man

Fig. 8. Five typical test images for the experiments.

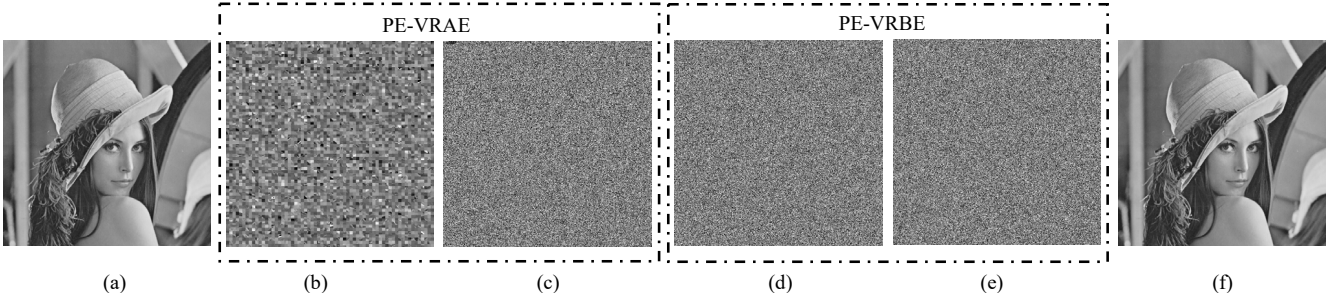

(a)  (b)  (c)  (d)  (e)  (f)

Fig. 9. Data embedding test. The original image cannot be directly observed from any of its encrypted versions. (a) Original image, (b) encrypted image of PE-VRAE (n = 8×8, ζ = 0.5), (c) marked encrypted image of PE-VRAE, (d) encrypted image of PE-VRBE, (e) marked encrypted image of PE-VRBE, (f) same error-free recovered image from (c) or (e) independently.

Table 3. The number of abnormal pixels with different parameters

| $n$ | 4×4 | | | | | 6×6 | | | | | 8×8 | | | | |
|---|---|---|---|---|---|---|---|---|---|---|---|---|---|---|---|
| $\zeta$ | 0.25 | 0.50 | 0.75 | 1.00 | None | 0.25 | 0.50 | 0.75 | 1.00 | None | 0.25 | 0.50 | 0.75 | 1.00 | None |
| Lena | 1 | 194 | 649 | 1399 | 6336 | 5 | 263 | 1035 | 2100 | 7576 | 0 | 310 | 1524 | 2895 | 9508 |
| Baboon | 5 | 23 | 93 | 524 | 16323 | 10 | 33 | 129 | 418 | 17654 | 8 | 31 | 125 | 385 | 19610 |
| Airplane | 6 | 187 | 962 | 1842 | 6693 | 7 | 199 | 1037 | 2024 | 8668 | 14 | 377 | 1262 | 2085 | 10081 |
| Tiffany | 15 | 33 | 48 | 121 | 5266 | 6 | 7 | 19 | 100 | 5910 | 26 | 34 | 39 | 99 | 7252 |
| Man | 561 | 1041 | 1951 | 3799 | 30213 | 880 | 1443 | 2625 | 4689 | 37349 | 1065 | 1734 | 3047 | 5339 | 42889 |

Table 4. The maximal net payload with different parameters by the proposed PE-VRAE scheme (*bpp*)

| $n$ | 4×4 | | | | | 6×6 | | | | | 8×8 | | | | |
|---|---|---|---|---|---|---|---|---|---|---|---|---|---|---|---|
| $\zeta$ | 0.25 | 0.50 | 0.75 | 1.00 | None | 0.25 | 0.50 | 0.75 | 1.00 | None | 0.25 | 0.50 | 0.75 | 1.00 | None |
| Lena | 3.107 | 3.103 | 3.094 | 3.082 | 2.952 | 3.229 | 3.224 | 3.210 | 3.198 | 3.065 | 3.322 | 3.314 | 3.293 | 3.285 | 3.143 |
| Baboon | 1.498 | 1.497 | 1.494 | 1.475 | 1.195 | 1.576 | 1.574 | 1.570 | 1.559 | 1.259 | 1.625 | 1.624 | 1.621 | 1.610 | 1.288 |
| Airplane | 3.383 | 3.378 | 3.364 | 3.355 | 3.256 | 3.535 | 3.531 | 3.516 | 3.503 | 3.398 | 3.637 | 3.630 | 3.620 | 3.611 | 3.495 |
| Tiffany | 3.235 | 3.235 | 3.234 | 3.234 | 3.098 | 3.359 | 3.359 | 3.358 | 3.358 | 3.219 | 3.441 | 3.441 | 3.441 | 3.440 | 3.292 |
| Man | 2.747 | 2.745 | 2.741 | 2.731 | 2.558 | 2.859 | 2.858 | 2.851 | 2.843 | 2.658 | 2.940 | 2.937 | 2.931 | 2.923 | 2.742 |
| Average | 2.794 | 2.792 | 2.785 | 2.775 | 2.612 | 2.912 | 2.909 | 2.901 | 2.892 | 2.720 | 2.993 | 2.989 | 2.981 | 2.974 | 2.792 |

### B. Performance Analysis

In Fig. 8, we select five examples of cover images to test the performances of the proposed schemes. We evaluate embedding capacity, security, separability, reversibility, and computational complexity. Fig. 9 shows some experiment results of the two proposed schemes on Lena.

***Embedding Capacity.*** Based on PE-VRAE, with different $n$ and ζ, the ER of five test images after image encryption is shown in Table 4. With a same block size, due to the influence of abnormal pixels in the precision of pixels prediction, the ER decreases gradually with the increasement of ζ. Especially, the average ER without using ζ is about 0.1*bpp* (bits per pixel) less than that with ζ = 1.00. With the same ζ, higher ER can be achieved by using larger block sizes. The reason is that the amount of reference pixels decreases with the increasement of block size, i.e., the amount of prediction errors increases. Table 4 indicates that the ER of Lena, Airplane and Tiffany can be above 3*bpp*, the ER of Man is up to 2.9*bpp*. Even for the textured image like Baboon, the ER can be 1.625*bpp* ($n = 8 \times 8$, $\zeta=0.25$).

Table 6. The average PSNR (dB) and SSIM of five test images with different parameters by the proposed PE-VRAE scheme

| Image | $n$ | 4×4 | | | | | 6×6 | | | | | 8×8 | | | | |
|---|---|---|---|---|---|---|---|---|---|---|---|---|---|---|---|---|
| | $\zeta$ | 0.25 | 0.50 | 0.75 | 1.00 | None | 0.25 | 0.50 | 0.75 | 1.00 | None | 0.25 | 0.50 | 0.75 | 1.00 | None |
| Encrypted image | PSNR | 13.25 | 13.70 | 13.54 | 12.90 | 9.38 | 13.28 | 13.66 | 13.52 | 12.94 | 9.41 | 13.03 | 13.38 | 13.27 | 12.79 | 9.33 |
| | SSIM | 0.305 | 0.390 | 0.366 | 0.280 | 0.073 | 0.357 | 0.434 | 0.430 | 0.352 | 0.086 | 0.336 | 0.395 | 0.388 | 0.323 | 0.090 |
| Marked encrypted image | PSNR | 9.48 | 9.53 | 9.53 | 9.50 | 9.30 | 9.45 | 9.46 | 9.47 | 9.44 | 9.36 | 9.39 | 9.40 | 9.40 | 9.40 | 9.32 |
| | SSIM | 0.046 | 0.045 | 0.044 | 0.045 | 0.043 | 0.043 | 0.044 | 0.044 | 0.041 | 0.043 | 0.042 | 0.041 | 0.044 | 0.041 | 0.042 |

Table 7. The average PSNR (dB) and SSIM of five test images by the proposed VRBE schemes

| | Encrypted image | | | | | | Marked encrypted image | | | | | |
|---|---|---|---|---|---|---|---|---|---|---|---|---|
| | Lena | Baboon | Airplane | Tiffany | Man | Average | Lena | Baboon | Airplane | Tiffany | Man | Average |
| PSNR | 9.23 | 9.52 | 8.01 | 6.88 | 7.99 | 8.33 | 9.23 | 9.51 | 8.01 | 6.90 | 8.00 | 8.33 |
| SSIM | 0.034 | 0.032 | 0.035 | 0.039 | 0.068 | 0.041 | 0.036 | 0.029 | 0.034 | 0.038 | 0.068 | 0.041 |

Table 5 shows the achieved ER based on the proposed PE-VRBE scheme. The average ER of the five test images is 3.116*bpp*, which proves that the proposed scheme can hide a large amount of additional data within a cover image. Compared with Table 4, Table 5 indicates that the achieved ER of PE-VRBE is higher than that of PE-VRAE. One reason is that PE-VRBE uses the spatial redundancy of the original plaintext image directly to vacate embedding room while the spatial redundancy is somewhat degraded during the image encryption in PE-VRAE, and another is that there is only one reference pixel in PE-VRBE while the number of reference pixels in PE-VRAE is equal to the number of image blocks.

Table 5. The maximal net payload based on PE-VRBE (*bpp*)

| Lena | Baboon | Airplane | Tiffany | Man | Average |
|---|---|---|---|---|---|
| 3.444 | 1.710 | 3.804 | 3.562 | 3.060 | 3.116 |

*Security.* We perform PE-VRAE on the five test images in Fig. 8 and Table 6 reports the average PSNR (dB) and SSIM of the encrypted images and the marked encrypted images towards the original images using different parameters. In PE-VRAE, the average PSNR and SSIM values between the original image and its encrypted version vary with different parameters $n$ and $\zeta$. Although there is blocking artifacts during block modulation and block permutation, the effect of the proposed encryption for VRAE is reasonable. Besides, there is no significant difference in average PSNR or SSIM with different $n$ and $\zeta$.

Table 7 shows the results of the proposed PE-VRBE scheme, where the PSNR and SSIM values between the original image and its encrypted version are very low. The average PSNR between the original image and the encrypted image with vacated room is 8.33dB, and the average SSIM is 0.041. For the marked encrypted image, the average PSNR is 8.33dB and the average SSIM is 0.041.

In both the PE-VARE and PE-VABE, the data embedding is imperceptible, since the marked encrypted image is garbled as the encrypted image. Even the receiver extracts the embedded data from the marked encrypted image, he cannot decrypt it to obtain the original additional data.

In PE-VRAE, without the encryption of the entropy encoded data, only the encrypted image can be recovered from the marked encrypted image. With one-time pad during image encryption, the proposed VRAE-based scheme can resist the known plaintext attacks presented in [24].

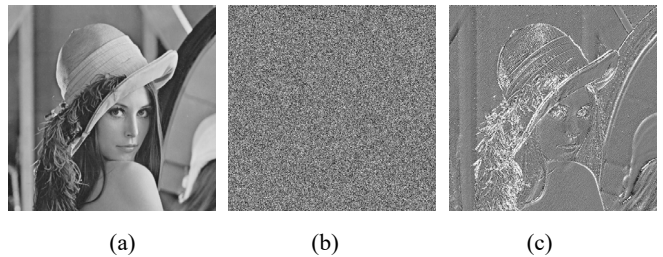

(a)      (b)      (c)

Fig. 10. The security flaw of method [33]. (a) Original image, (b) marked encrypted image, (c) contour leakage image.

As a comparison, in [33], any potential attacker can extract prediction errors from the encrypted image to get the content of the cover image. As shown in Fig. 10, (a) is the original image Lena sized 512×512; (b) is the marked encrypted image of (a) with RDHEI method in [33], where $\alpha = 5$, $\beta = 2$ and the ER is 2.645*bpp*; (c) is the resulted image obtained from the PBTL of (b), where the detected prediction errors in [−11, 12] are mapped to [0,255] for better visibility as an image. From Fig. 10 (c), the contour lines of (a) are clearly visible, the marked encrypted image can be identified in [33]. In contrast, in PE-VRBE, the data embedding in the encryption domain does not influence the encryption effect. Since the entropy encoded data is encrypted during image encryption, the contour lines of the original image can no longer be detected from the encrypted versions. Therefore, after image encryption and data embedding, little information of the cover image can be discovered from the marked encrypted version. As a result, while [33] has contour leakage issue, our scheme ensures better security and effectively prevents information leakage.

*Separability & Reversibility.* In the proposed schemes, the receiver gets different contents according to his authentication. The receiver only with the data hiding key can extract the hidden data exactly without knowing any information about the original content, and the data extraction is always successful, i.e., the bit-wise error rate (BER) between the extracted decrypted data and the original additional data is 0. Similarly, the receiver with only the encryption keys cannot obtain the hidden data, but he can recover the original image losslessly. The recovered image is identical to the original image, e.g., Fig. 9 (f) is identical to Fig. 9 (a). The above experiments demonstrate the separability and reversibility of the proposed schemes.

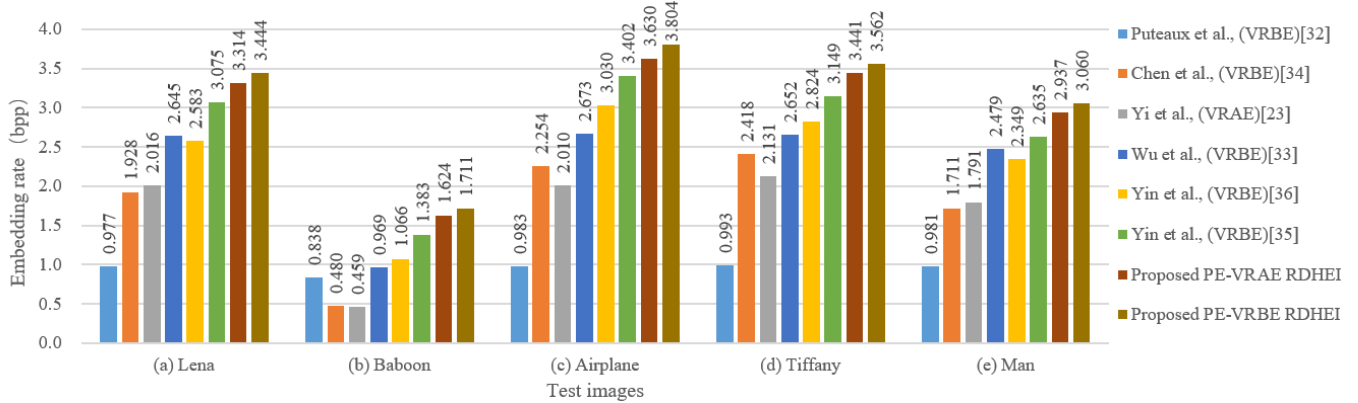

Fig. 11. Comparison of ER on five test images between the proposed schemes and six state-of-the-art RDHEI methods.

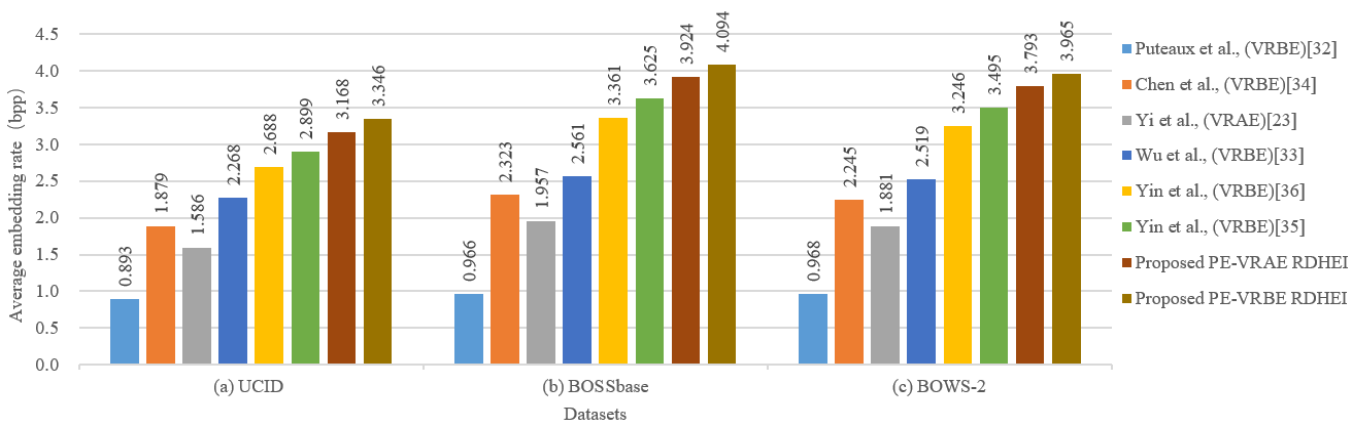

Fig. 12. Comparison of average ER on three image datasets between the proposed schemes and six state-of-the-art RDHEI methods.

***Computational complexity.*** The computational complexity of the proposed schemes mainly depends on the entropy encoding. We evaluate the runtime of the arithmetic coding in the proposed PE-VRAE ($n = 8\times 8$, $\zeta =0.50$) and PE-VRBE schemes. The configurations of our PC are Intel Core i5-10210U CPU, 8 GB RAM, Windows 10 64-bit operating system, and the software is MATLAB R2016a. The running times of the arithmetic encoding on five test images are listed in Table 8. Since the size of Man is 1024×1024 and that of other images is 512×512, the runtime on Man is larger than that of other images. The runtime of arithmetic encoding is reasonable, and there is no significant different between PE-VRAE and PE-VRBE.

Table 8. The runtime of arithmetic encoding in proposed schemes(s)

|  | Lena | Baboon | Airplane | Tiffany | Man | AVG |
|---|---|---|---|---|---|---|
| PE-VRAE | 0.280 | 0.504 | 0.283 | 0.271 | 1.435 | 0.555 |
| PE-VRBE | 0.241 | 0.441 | 0.249 | 0.244 | 1.803 | 0.596 |

### C. Performance Comparison with State-of-the-art RDHEI Schemes

We compare the proposed schemes with several state-of-the-art RDHEI schemes [23] [32-36] in terms of ER when the PSNR and SSIM of the marked encrypted image produced by the schemes are close. A larger ER means a better embedding performance. In the experimental settings, the parameters α and β in [23] and [33] are set as 5 and 2 respectively, the block size in [23] is 3×3. In [32], we accept the EPE-HCRDH scheme, which can achieve a higher embedding capacity. In [34], the length of fixed-length codewords is set as 3 and the block size is set to be 4×4. In [35], the 8 bit-plants of prediction errors are rearranged and compressed. In [36], the codewords are predefined as {00, 01, 100, 101, 1100, 1101, 1110, 11110, 11111} to represent nine kinds of the multi-MSB prediction. In the proposed PE-VRAE scheme, we set $n = 8\times 8$ and $\zeta =0.50$.

***Embedding Rate***. Fig. 11 provides the performance comparison of embedding rate between the two proposed schemes and methods in [23] [32-36]. The embeddings are conducted on the cover images shown in Fig. 8. Although the payloads of previous VRAE-based schemes are reported to be naturally lower than those of VRBE-based schemes, we find that our PE-VRAE can beat some state-of-the-art VRBE-based schemes, owing to the effectiveness of the correlation preserving and prediction accuracy. Besides, the ER of the proposed PE-VRAE scheme is close to that of the proposed PE-VRBE scheme. The achieved ER of EPE-HCRDH [32] is lower than 1*bpp*. The reason is that the method merely uses the MSB of each pixel for data embedding. Chen et al. [34] can provide a higher embedding capacity. However, the method performs poorly on textured images, e.g., 0.459*bpp* on Baboon. Yi et al. [23] provides a similar embedding performance compared to [34], where the net payload on Baboon is 0.480*bpp*. Methods in [33] [35-36] can achieve higher payloads on both flat and textured images. For example, the method in [35] can achieve embedding rate of 3.075*bpp*, 1.383*bpp*, 3.402*bpp*, 3.149*bpp* and 2.635*bpp* on Lena, Baboon, Airplane, Tiffany, and Man respectively. By comparison, the proposed schemes outperform above methods in net payload. With the proposed VRAE-based method by arithmetic coding, the ER is up to 3.314*bpp*, 1.624*bpp*, 3.630*bpp*, 3.441*bpp* and 2.937*bpp* on Lena, Lena, Baboon, Airplane, Tiffany, and Man respectively. While with the proposed VRBE-based method, the ER is up to 3.444*bpp*, 1.711*bpp*, 3.804*bpp*, 3.562*bpp* and 3.060*bpp* in five test images respectively. The reason is that the proposed schemes not only take full advantage of the spatial redundancy in the plain-text image, but also improve the performance of entropy coding with arithmetic coding.

Table 9. Detailed Embedding Rate Comparisons on Textured Images

| Size | Method | 1 | 2 | 3 | 4 | 5 | 6 | 7 | 8 | 9 | 10 | 11 | 12 | 13 | Average |
|---|---|---|---|---|---|---|---|---|---|---|---|---|---|---|---|
| 512×512 | [33] | 0.119 | 1.094 | 0.395 | 1.153 | 1.046 | 0.375 | 1.446 | 2.937 | 2.758 | 2.155 | 1.807 | 1.608 | 1.538 | 1.418 |
| | [35] | 0.809 | 1.475 | 1.132 | 1.450 | 1.549 | 0.928 | 1.614 | 3.356 | 3.017 | 2.220 | 2.062 | 1.830 | 1.750 | 1.784 |
| | [36] | 0.664 | 1.230 | 0.812 | 1.316 | 1.247 | 0.888 | 1.537 | 2.956 | 2.617 | 1.951 | 1.829 | 1.668 | 1.495 | 1.554 |
| | **PE-VRBE** | **1.256** | **1.984** | **1.529** | **2.023** | **2.026** | **1.462** | **2.172** | **3.853** | **3.538** | **2.819** | **2.564** | **2.326** | **2.236** | **2.291** |
| | PE-VRAE | 1.104 | 1.794 | 1.412 | 1.878 | 1.944 | 1.340 | 2.042 | 3.570 | 3.168 | 2.552 | 2.423 | 2.197 | 2.092 | 2.117 |
| 1024×1024 | [33] | 1.360 | 1.574 | 1.813 | 0.730 | 0.685 | 1.410 | 2.029 | 2.143 | 2.478 | 2.304 | 1.916 | 1.877 | 1.786 | 1.700 |
| | [35] | 1.790 | 1.934 | 2.126 | 1.258 | 1.405 | 1.809 | 2.269 | 2.375 | 2.603 | 2.466 | 2.193 | 2.136 | 2.101 | 2.036 |
| | [36] | 1.465 | 1.582 | 1.735 | 1.053 | 1.098 | 1.413 | 1.873 | 1.997 | 2.396 | 2.053 | 1.835 | 1.861 | 1.688 | 1.696 |
| | **PE-VRBE** | **2.274** | **2.415** | **2.578** | **1.717** | **1.841** | **2.311** | **2.712** | **2.822** | **3.062** | **2.920** | **2.659** | **2.576** | **2.555** | **2.496** |
| | PE-VRAE | 2.097 | 2.282 | 2.433 | 1.621 | 1.682 | 2.146 | 2.606 | 2.772 | 2.932 | 2.824 | 2.565 | 2.469 | 2.455 | 2.376 |

Apart from the five typical images, we further analyze the average embedding performance on UCID [38], BOSSBase [39], and BOWS-2 [40]. Fig. 12 reports the performances of average embedding rate on the three datasets. While the method in [35] can embed a larger payload into the cover image than methods in [23, 32-34, 36], the proposed schemes can further offer a larger net payload. The experiment results demonstrate that the proposed schemes outperform other competitors. For example, the average net payload of the proposed VRBE-based scheme on UCID, BOSSbase 1.01 and BOWS-2 is 3.346*bpp*, 4.094*bpp* and 3.965*bpp*, which is 15.41%, 12.94% and 13.45% larger than that of [35], respectively. The average net payloads of the proposed PE-VRAE scheme on UCID, BOSSbase 1.01 and BOWS-2 is 3.168*bpp*, 3.924*bpp* and 3.793*bpp*, which is respectively 68.60%, 68.92% and 68.95% larger than that of the VRAE-based method [23].

Besides, with the two existing RDHEI methods based on pixel prediction and entropy encoding of prediction errors, i.e., VRAE-based method [23] and VRBE-based method [33], the ER of [33] is much higher than that of [23], especially the ER on texture image Baboon is relatively low. However, based on pixel prediction and entropy encoding under the proposed framework, the proposed PE-VRAE scheme is close to that of the proposed PE-VRBE scheme, and the ERs are respectively up to 1.624*bpp* and 1.711*bpp* on Baboon.

*Average performance on textured images.* We validate the proposed schemes on textured images using USC-SIPI image database [42]. We conduct experiments on two resolutions 512×512 and 1024×1024 of 13 different images. As showed in Table 9, we also compare the embedding rate of the proposed schemes with methods in [33, 35, 36]. It is obvious that the proposed schemes achieve higher net payload than [33, 35, 36] in all the 26 textured images sized 512×512 or 1024×1024. Besides, the achieved ERs of 1024×1024 images are higher than that of 512×512 images. The experiment further demonstrates that the proposed schemes can achieved higher embedding capacity on the textured images.

## VI. CONCLUSION

This paper presents a general framework for high-capacity RDHEI by using pixel prediction and entropy encoding, which can be applied to both VRAE and VRBE cases. In the proposed framework, an efficient embedding room generation algorithm is designed to produce large embedding room by entropy encoding of pixel prediction errors. We then propose two RDHEI schemes, denoted as PE-VRBE and PE-VRAE respectively. In PE-VRBE, the image owner generates embedding room using ERGA and encrypts the preprocessed image by the stream cipher. Then, the data hider locates the embedding room and embeds the encrypted additional data to the encrypted image. In PE-VRAE, the cover image is encrypted by an improved block modulation and permutation encryption algorithm to preserve the spatial redundancy in the plain-text image. By applying ERGA, the data hider generates the embedding room from the encrypted image and conducts data embedding to obtain the marked encrypted image. For both the schemes, the receivers with different authentication keys can respectively conduct error-free data extraction and/or error-free image recovery. Experimental results prove the effectiveness of the proposed schemes on typical image datasets and the proposed schemes outperform many state-of-the-art RDHEI works. Besides, while many previous works cannot provide large embedding rate on textured images like Baboon, the proposed PE-VRBE scheme can provide the data hider with up to 1.711$bpp$ payload for Baboon.

APPENDIX. EXTENSION WITH HUFFMAN ENCODING

In this appendix, we replace the arithmetic encoding used in the proposed ERGA with Huffman encoding to verify the adaptiveness of the proposed framework. Huffman code is another common entropy encoding paradigm. Differing from arithmetic coding which encodes all embedding pixels into a single number, Huffman encoding replaces each independent encoding pixel with an independent codeword and all joint encoding pixels with a same codeword.

For a given $T \in [1, 255]$, we build an optimal Huffman tree $\mathfrak{h}$ with codewords $\{C_{-T}, C_{-T+1}, …, C_{-1}, C_0, …, C_{T-2}, C_{T-1}, C_J\}$. We assign each prediction error in the independent encoding bins with an independent codeword, and the joint encoding bins with a same codeword. The codewords of the independent encoding pixels are $\{C_{-T}, C_{-T+1}, …, C_{-1}, C_0, …, C_{T-2}, C_{T-1}\}$ with $\ell(C_e)$ ($e \in [-T, T-1]$) bits, and that of the joint encoding pixels is $C_J$ with $\ell(C_J)$ bits.

For the image decompression, we record the information for constructing the Huffman tree, including the threshold $T$, the (2$T$+1) codewords and the length of each codeword. For $T \in [1, 255]$, it can be expressed within eight bits. Since the maximum length of each pixel is less than eight, the length of each codeword should be less than eight so that every embedding pixel can accommodate its codeword to vacate embedding room. The length of each codeword can be represented within three bits, appended with its corresponding binary codeword. The above side information is concatenated as a binary sequence with length $\ell_T(CD_1)$ as

$$\ell_T(CD_1) = 8+3(2T+1) + \sum_{e=-T}^{T-1}\ell(C_e) + \ell(C_J). \quad (30)$$

For each *embedding pixel* $y_{i,j,k}$ ($1 \leq i \leq N, j \neq j_1$ or $k \neq k_1$), its prediction error $e_{i,j,k}$ can be represented by its corresponding Huffman codeword $c_{i,j,k}$. When $e_{i,j,k} \in [-T, T-1]$, the codeword $c_{i,j,k}$ is $C_{e_{i,j,k}}$, otherwise the codeword $c_{i,j,k}$ is $C_J$. Given a $T$, total length of codewords used for recording the prediction errors is defined in Eq. (31).

$$\ell_T(CD_2) = \sum_{e=-T}^{T-1} h_e \cdot \ell(C_e) + h_J \cdot \ell(C_J). \quad (31)$$

Then, $CD_1$ is appended with $CD_2$, we have

$$\ell_T(CD) = \ell_T(CD_1) + \ell_T(CD_2) = 8+3(2T+1) + \ell_T(CD'), (32)$$

where,

$$\ell_T(CD') = [\sum_{e=-T}^{T-1}(h_e + 1)\ell(C_e) + (h_J + 1)\ell(C_J)]. \quad (33)$$

With Huffman encoding, the embedding capacity of vacated room in Eq. (9) can be described as

$$EC_T = 8 \cdot (\sum_{e=-T}^{T-1} h_e - 1) - 3(2T+1) - \ell(L) - \ell_T(CD'). \quad (34)$$

Then, to achieve maximal embedding capacity, the optimal threshold $T_{opt}$ should satisfy

$$T_{opt} = \underset{T \in [1,255]}{\text{argmax}} EC_T,$$

w.r.t. $\forall \ell(C_e) \leq 8, \ell(C_J) \leq 8, e \in [-T, T-1]$. (35)

For a special threshold $T$, the value of the first three items in Eq. (34) is fixed, the optimal Huffman tree should be constructed to minimize the value of $\ell(CD')$, i.e.,

$$\mathfrak{h}_T = \text{argmin } \ell(CD'),$$

w.r.t. $\forall \ell(C_e) \leq 8, \ell(C_N) \leq 8, e \in [-T, T-1]$. (36)

For each prediction error value $e \in [-T, T-1]$, the occurrence probability of its codeword $C_e$ is

$$p(C_e) = (h_e + 1)/(\sum_{e'=-255}^{255} h_{e'} + 2 \cdot T + 1), \quad (37)$$

where,

$$\sum_{e'=-255}^{255} h_{e'} = N \cdot (n_1 \cdot n_2 - 1). \quad (38)$$

The total occurrence probability of codeword $C_J$ for all $P_J$ is

$$p(C_J) = 1 - \sum_{e=-T}^{T-1} p(C_e). \quad (39)$$

Based on the probability distribution, a Huffman tree can be generated for a special $T$. By trying all possible $T$, the optimal threshold $T_{opt}$ can be determined with Eq. (35). In this way, the $T_{opt}$ and its corresponding optimized Huffman tree are generated, and the prediction errors can be compressed with Huffman encoding to vacate embedding room.

In PE-VRAE ($n = 8 \times 8$, $\zeta$=0.25), the achieved ER of Lena, Baboon, Airplane, Tiffany, and Man by Huffman encoding is 3.187$bpp$, 1.454$bpp$, 3.497$bpp$, 3.306$bpp$ and 2.817$bpp$ respectively, while the average achieved ER on UCID [38], BOSSBase [39], and BOWS-2 [40] by Huffman encoding is respectively 3.043$bpp$, 3.625$bpp$ and 3.669$bpp$. These experimental data is a bit lower than that of PE-VRAE with arithmetic coding, but higher than that of [23] [32-36]. With PE-VRBE by Huffman encoding, the achieved ER of Lena, Baboon, Airplane, Tiffany, and Man is 3.317$bpp$, 1.535$bpp$, 3.657$bpp$, 3.427$bpp$ and 2.944$bpp$ respectively, the average achieved ER on UCID [38], BOSSBase [39], and BOWS-2 [40] is 3.217$bpp$, 3.952$bpp$ and 3.814$bpp$ respectively. These experimental data is a bit lower than that of PE-VRBE with arithmetic coding, but also higher than that of [23] [32-36]. These results indicate that the Huffman code is also fit to the proposed framework of RDHEI. In addition, applying different entropy encoding methods has little effect on the security, separability, and reversibility of the proposed RDHEI schemes.